
\input amstex
\documentstyle{amsppt}
\magnification=1200
\leftheadtext{D.Juriev}
\rightheadtext{Noncommutative Veronese Mapping}
\NoBlackBoxes
  \define\CA{\Cal A}

  \define\CF{\Cal F}
\define\Fg{\frak g}

\define\Fl{\frak l}

\define\Fs{\frak s}  \define\CS{\Cal S}
  
  \define\CU{\Cal U}


\define\AC{\Bbb C}   \define\BC{\bold C}

\define\AQ{\Bbb Q}   
\define\AR{\Bbb R}

\define\AZ{\Bbb Z}   
\define\Tl{\text{\tt l}}
\define\Tw{\text{\tt w}}
 \define\vtheta\vartheta
 
\define\ga{\alpha}

\define\gd{\delta}


\define\End{\operatorname{End}}

\define\Der{\operatorname{Der}}

\define\ad{\operatorname{ad}}

\define\pd#1#2{\partial_{#1}^{#2}}
\define\p#1{\partial_{#1}}
\define\fpd#1#2{(\frac{\partial}{\partial_{#1}})^{#2}}
\define\fp#1{\frac{\partial}{\partial_{#1}}}
\define\Gr#1#2#3{#1(#2,#3)}
\define\GrC#1#2{\Gr#1#2{\AC}}
\define\GrR#1#2{\Gr#1#2{\AR}}
\define\GrQ#1#2{\Gr#1#2{\AQ}}
\define\GrZ#1#2{\Gr#1#2{\AZ}}
\define\vGr#1#2#3{#1_{#2}(#3)}
\define\vGrC#1#2{\vGr#1#2{\AC}}
\define\vGrR#1#2{\vGr#1#2{\AR}}
\define\vGrQ#1#2{\vGr#1#2{\AQ}}
\define\vGrZ#1#2{\vGr#1#2{\AZ}}

 
\define\sla{\operatorname{\Fs\Fl}}

\define\sLtwo{\GrC{\sla}2}
\define\sLthree{\GrC{\sla}3}
\define\RW{\operatorname{\Cal R\Cal W}}
\define\cRW{\widehat{\RW}}
\define\RWtwo{\RW(\sLtwo)}
\define\cRWtwo{\cRW(\sLtwo)}
\define\RWtwop#1{\RW(\sLtwo;#1)}
\define\cRWtwop#1{\cRW(\sLtwo;#1)}
\define\pz{\p z}
\define\pzs{\pd z 2}

\topmatter
\title SETTING HIDDEN SYMMETRIES FREE BY THE NONCOMMUTATIVE VERONESE MAPPING
\endtitle
\author\eightpoint D.JURIEV\footnote{On leave from Mathematical Division,
Research Institute for System Studies, Russian Academy of Sciences, Moscow,
Russia (e-mail: juriev\@systud.msk.su)\newline}
\centerline{}
\centerline{}
\centerline{}
\centerline{Laboratoire de Physique Th\'eorique de l'\'Ecole Normale
Sup\'erieure,}
24 rue Lhomond, 75231 Paris Cedex 05, France\footnote{Unit\'e Propre du Centre
National de la Recherche Scientifique associ\'ee \`a l'\'Ecole Normale
Sup\'erieure et \`a l'Universit\'e de Paris--Sud\newline}
\centerline{E mail: juriev\@physique.ens.fr}
\endauthor
\abstract The note is devoted to the setting free of hidden symmetries in
Verma modules over $\sLtwo$ by the noncommutative Veronese mappings.
\endabstract
\endtopmatter
\document

In many cases the behavior of systems is governed not only by their natural
(geometric) symmetries but also by hidden ones. The main difficulty to work
with hidden symmetries is that they are often "packed", and as a rule can't be
"unpacked" to the universal envelopping algebras of Lie algebras, so there
exists a problem how to set them free "correctly" (see f.e.[1-3]). That means
to find a "correct" algebraic structure, which is represented by them. This is
one of the themes of this short note. It maybe considered as preliminary to
the second one, which is related to a view of the noncommutative geometry [1]
on the setting hidden symmetries free. The most interesting setting free
mappings are noncommutative Veronese mappings (certain analogs of classical
ones [4]);  on such way the problem of the noncommutative "birational"
equivalence of the differently obtained set free hidden symmetries is appeared
(cf.[5]).  Both marked subjects are interacted in the paper on the simplest
examples of hidden symmetries in Verma modules over $\sLtwo$.

I wish to thank Prof. J.-L.Gervais and the collective of LPTENS for a kind
hospitality and sincere atmosphere during my work on the note.

\definition{Definition}

{\bf A.} Let $\Fg$ be a Lie algebra and $\CA$ be an associative algebra such
that $\Fg\subset\Der(\CA)$; a linear subspace $V$ of $\CA$ is called {\it a
space of hidden symmetries\/} iff (1) $V$ is a $\Fg$--submodule of $\CA$, (2)
the Weyl symmetrization defines a surjection $W:S^{\cdot}(V)\mapsto\CA$ (the
elements of $V$ are called {\it hidden symmetries with respect to $\Fg$}). An
associative algebra $\CF$ such that $\Fg\subset\Der(\CF)$ is called {\it an
algebra of the set free hidden symmetries\/} iff (1) $\CF$ is generated by
$V$, (2) there exists a $\Fg$--equivariant epimorphism of algebras
$\CF\mapsto\CA$, (3) the Weyl symmetrization defines an isomorphism
$S^{\cdot}(V)\mapsto\CF$ of $\Fg$--modules.

{\bf B.} Let $V$ be a space of hidden symmetries and $\BC$ is some class of
associative Ore algebras; the hidden symmetries from $V$ are called {\it
$\BC$--regular\/} iff the algebras of quotients $D(\CF)$ are isomorphic for
all corresponding to $V$ algebras $\CF$ of the set free hidden symmetries from
the class $\BC$.

{\bf C.} Let $V$ be a space of hidden symmetries in algebra $\CA$ with respect
to the Lie algebra $\Fg$; a subspace $V_0$ of $V$ is called {\it a coordinate
base\/} of $V$ iff (1) $V_0$ is a $\Fg$--submodule of $V$, (2) the image of
the Weyl symmetrization mapping $W_0:S^{\cdot}(V_0)\mapsto\CA$ contains $V$.
Hidden symmetries from $V$ are called {\it of type $(V_0,n)$\/} iff the image
of
$\bigoplus_{i\le n}S^i(V_0)$ under the Weyl symmetrization mapping $W_0$
coincides with $V$; in this case the mapping $\bigoplus_{i\le
n}S^i(V_0)\mapsto\CF$, a composition of $W_0$ and the imbedding of $V$ into
$\CF$, is called {\it the noncommutative Veronese mapping}.

{\bf D.} Let $\Fg$ be a Lie algebra, $V$ be a certain $\Fg$--module, $\CA_s$
be a family of associative algebras, parametrized by $s\in\CS$ such that
$\Fg\subset\Der(\CA_s)$, $\pi_s:V\mapsto\CA_s$ be a family of
$\Fg$--equivariant imbeddings such that $\pi_s(V)$ is a space of hidden
symmetries in $\CA_s$ with respect to $\Fg$ for a generic $s$ from $\CS$.  An
associative algebra $\CF$ is called {\it an algebra of the
$\CA_{s,s\in\CS}$--universally set free hidden symmetries\/} iff $\CF$ is an
algebra of the set free hidden symmetries corresponding to $V\simeq\pi_s(V)$
for generic $\CA_s$ ($s\in\CS$). The hidden symmetries are called {\it the
$\CA_{s,s\in\CS}$--universally $\BC$--regular\/} iff the algebras of quotients
$D(\CF)$ are isomorphic for all algebras $\CF$ of the
$\CA_{s,s\in\CS}$--universally set free hidden symmetries from the class
$\BC$.
\enddefinition

If $\Fg$ is a Lie algebra and $\CA$ is an associative algebra such that
$\Fg\subset\Der(\CA)$, $V_0$ is a $\Fg$--submodule of $\CA$, which elements
generate $\CA$ as an algebra then in many interesting cases there exists a
space of hidden symmetries $V$ of type $(V_0,n)$ in $\CA$ for a sufficiently
large $n$.

\proclaim{Theorem}

{\bf A.} The tensor operators of type $\pi_1$ and $\pi_2$ in the Verma module
$V_h$ over the Lie algebra $\sLtwo$ ($\pi_i$ is a finite--dimensional
representation of $\sLtwo$ of dimension $2i+1$) form a space of hidden
symmetries of type $(\pi_1,2)$;
the quadratic (non--homogeneous)
algebras of the $\End(V_h)$--universally set free hidden symmetries form an
one--parametric family
$\RWtwop\ga$, where $\RWtwop 0 $ is the Racah--Wigner algebra
$\RWtwo$ of par.2.2. of ref [6].

{\bf B.} All the algebras of quotients $D(\RWtwop\ga)$ are isomorphic
(hence, the tensor operators of type $\pi_i$ ($i=1,2$) form a
$\End(V_h)$--universally quadratic--regular scope of hidden symmetries).

{\bf C.} The central extension $\cRWtwo$ of $\RWtwo$ maybe continued to the
central extensions $\cRWtwop\ga$ of $\RWtwop\ga$ in the class of
quadratic (non--homogeneous) algebras.
\endproclaim

{\bf Comments on Th.1A,1C.} The statements 1A and 1C are verified by
explicit calculations. Here we present the constructions of algebras $\RWtwop
\ga$ and $\cRWtwop \ga$.

Let $L_i$ be a basis in $\sLtwo$ such that $[L_i,L_j]=(i-j)L_{i+j}$ and
$d^k_j$ ($-k\le j\le k$) be basises in $\pi_k$, in which the $\sLtwo$--action
has the form $L_i(d^k_j)=(ki-j)d^k_{i+j}$. The corresponding tensor operators
in the Verma modules $V_h$ will be denoted by the capitals. If the Verma
module $V_h$ is realized in the space $\AC[z]$ of polynomials of a complex
variable $z$, where the generators of $\sLtwo$ act as $L_{-1}=z$,
$L_0=z\pz+h$ and $L_1=z(\pz)^2+2h\pz$, then the tensor
operators $D^k_i$ ($k=1,2,3$) are defined by the formulas
$$\allowdisplaybreaks\align
D_{-1}^1&=z\\
D_0^1&=\xi+h\\
D_1^1&=(\xi+2h)\pz\\
     & \\
D_{-2}^2&=z^2\\
D_{-1}^2&=z(\xi+h+\tfrac12)\\
D_0^2&=\xi^2+2h\xi+\tfrac{h(2h+1)}3\\
D_1^2&=(\xi+2h)(\xi+h+\tfrac12)\pz\\
D_2^2&=(\xi+2h)(\xi+2h+1)\pzs
\endalign
$$
where $\xi=z\pz$.

Algebra $\RWtwop\ga$ is generated by $l_i$ ($i=-1,0,1$) and $w_i$
($i=-2,-1,0,1,2$); the action of $\sLtwo$ has the form
$L_i(l_j)=(i-j)l_{i+j}$, $L_i(w_j)=(2i-j)w_{i+j}$;
the commutation relations are following
$$\allowdisplaybreaks\align
[l_{-1},l_0]&=-l_{-1}+\ga(l_1\circ w_{-2}-2l_0\circ w_{-1}+l_{-1}\circ w_0)\\
[l_{-1},l_1]&=-2l_0+2\ga(l_1\circ w_{-1}-2l_0\circ w_0+l_{-1}\circ w_1)\\
[l_0,l_1]&=-l_1+\ga(l_1\circ w_0-2l_0\circ w_1+l_{-1}\circ w_2)\\
    &     \\
[l_{-1},w_{-1}]&=-w_{-2}+4\ga l_{-1}^2+4\ga(w_{-2}\circ w_0-w_{-1}^2)\\
[l_{-1},w_0]&=-2w_{-1}+8\ga l_{-1}\circ l_0+
4\ga(w_{-2}\circ w_1-w_{-1}\circ w_0)\\
[l_{-1},w_1]&=-3w_0+4\ga(l_{-1}\circ l_1+2l^2_0)+
2\ga(w_{-2}\circ w_2+2w_{-1}\circ w_1-3w^2_0)\\
[l_{-1},w_2]&=-4w_1+16\ga l_0\circ l_1+
8\ga(w_{-1}\circ w_2-w_0\circ w_1)\\
[l_0,w_{-2}]&=2w_{-2}-8\ga l^2_{-1}-8\ga(w_{-2}\circ w_0-w^2_{-1})\\
[l_0,w_{-1}]&=w_{-1}-4\ga l_{-1}\circ l_0-
2\ga(w_{-2}\circ w_1-w_{-1}\circ w_0)\\
[l_0,w_0]&=0\\
[l_0,w_1]&=-w_1+4\ga l_0\circ l_1+
2\ga(w_{-1}\circ w_2-w_0\circ w_1)\\
[l_0,w_2]&=-2w_2+8\ga l^2_1+8\ga(w_0\circ w_2-w^2_1)\\
[l_1,w_{-2}]&=4w_{-1}-16\ga l_{-1}\circ l_0-
8\ga(w_{-2}\circ w_1-w_{-1}\circ w_0)\\
[l_1,w_{-1}]&=3w_0-4\ga(l_{-1}\circ l_1+2l^2_0)-
2\ga(w_{-2}\circ w_2+2w_{-1}\circ w_1-3w^2_0)\\
[l_1,w_0]&=2w_1-8\ga l_0\circ l_1-
4\ga(w_{-1}\circ w_2-w_0\circ w_1)\\
[l_1,w_1]&=w_2-4\ga l^2_1-4\ga(w_0\circ w_2-w^2_1)\\
    &     \\
[w_{-2},w_{-1}]&=-2l_{-1}\circ w_{-2} \\
[w_{-2},w_0]&=-\tfrac43(2l_{-1}\circ w_{-1}+l_0\circ w_{-2}) \\
[w_{-2},w_1]&=-3l_{-1}\circ w_0-2l_0\circ w_{-1}-l_1\circ w_{-2}\\
[w_{-2},w_2]&=-4(l_{-1}\circ w_1+l_1\circ w_{-1})\\
[w_{-1},w_0]&=-\tfrac16(3l_{-1}\circ w_0+10 l_0\circ w_{-1}-l_1\circ w_{-2})\\
[w_{-1},w_1]&=-\tfrac12(l_{-1}\circ w_1+6l_0\circ w_0+l_1\circ w_{-1}) \\
[w_{-1},w_2]&=-l_{-1}\circ w_2-2l_0\circ w_1-3l_1\circ w_0\\
[w_0,w_1]&=\tfrac16(l_{-1}\circ w_2-10 l_0\circ w_1-3l_1\circ w_0)\\
[w_0,w_2]&=-\tfrac43(2l_1\circ w_1+l_0\circ w_2) \\
[w_1,w_2]&=-2l_1\circ w_2,
\endalign
$$
where $a\circ b=\frac{ab+ba}2$.

The algebra $\RWtwop\ga$ admits a representation by tensor operators in
the Verma module $V_h$ over $\sLtwo$ by $l_i\mapsto\gd^{-1}D_i^1$,
$w_i\mapsto\gd^{-1}D_i^2$ ($\gd=1-\frac{(2h+1)(2h+3)}3\ga$).

The algebra $\cRWtwop\ga$ is generated by $l_i$ ($i=-1,0,1$), $w_i$
($i=-2,-1,0,1,2$) and the central element $\rho$. The commutation relations
coincide with ones for $\RWtwop\ga$ up to subsidiary terms for $[w_i,w_j]$.
Namely, the improved commutators have the form
$$\allowdisplaybreaks\align
[w_{-2},w_{-1}]&=-2l_{-1}\circ w_{-2} \\
[w_{-2},w_0]&=-\tfrac43(2l_{-1}\circ w_{-1}+l_0\circ w_{-2}) \\
[w_{-2},w_1]&=-3l_{-1}\circ w_0-2l_0\circ w_{-1}-l_1\circ w_{-2}-\rho l_{-1}\\
[w_{-2},w_2]&=-4(l_{-1}\circ w_1+l_1\circ w_{-1}+\rho l_0)\\
[w_{-1},w_0]&=-\tfrac16(3l_{-1}\circ w_0+10 l_0\circ w_{-1}-l_1\circ
w_{-2}+3\rho l_{-1})\\
[w_{-1},w_1]&=-\tfrac12(l_{-1}\circ w_1+6l_0\circ w_0+l_1\circ w_{-1}+\rho l_0)
\\
[w_{-1},w_2]&=-l_{-1}\circ w_2-2l_0\circ w_1-3l_1\circ w_0-\rho l_1\\
[w_0,w_1]&=\tfrac16(l_{-1}\circ w_2-10 l_0\circ w_1-3l_1\circ w_0+3\rho l_1)\\
[w_0,w_2]&=-\tfrac43(2l_1\circ w_1+l_0\circ w_2) \\
[w_1,w_2]&=-2l_1\circ w_2
\endalign
$$

{\bf Sketch of the proof of Th.1B.} The statement 1B is proved by an explicit
construction of isomorphism in 4 steps. 1st step: operators $L_i$ are
represented in the form $L_i=\ad(\Tl_i)=\ad(l_i)+\sum_{j\ge
1}\ga^j\ad(X_{i,j})$, $X_{i,j}\in\RWtwop \ga$. 2nd step: there exists the
unique operators $\Tw_i=w_i+\sum_{j\ge 1}\ga^j Y_{i,j}$, $Y_{i,j}\in\RWtwop
\ga$ such that $[\Tl_i,\Tw_j]=(2i-j)\Tw_{i+j}$. 3rd step: $\Tl_i$ and $\Tw_j$
obey the commutation relations of $\RWtwo$. By these three steps we proved
that the deformation $\{\RWtwop\ga; \ga\in\AC\}$ of $\RWtwo$ is formally
trivial. 4th step: formal series for $\Tl_i$ and $\Tw_j$ are rational.

{\bf Remark.} $\RWtwo$ is a Hopf algebra [1,7] with the co--commutative
comultiplication $l_i\mapsto l_i\otimes 1+1\otimes l_i$, $w_i\mapsto
w_i\otimes 1+1\otimes w_i+c^{jk}_i l_j\otimes l_k$, where
$c=\{c^{jk}_i\}\in[\pi_2\otimes S^2(\pi_1)]^{\sLtwo}$ and the antipode
$1\mapsto 1$, $l_i\mapsto -l_i$, $w_j\mapsto w_j$. The structure of Hopf
algebra on $\RWtwo$ is a deformation of such structure on $\CU(\sLthree)$
(with non--standard antipode corresponding to the Cartan antiautomorphism).
This deformation is realized by the fixing of structures of Hopf algebras on
$\cRWtwo/(\rho\!\!=\!\!\rho_0)$
($\cRWtwo/(\rho\!\!=\!\!\rho_0)\longrightarrow_{\sssize{\rho_0\to\infty}}\CU(\sLthree)$
par.2.2 of ref [6]). Unfortunately, I do not know a way to make a Hopf algebra
from $\RWtwop \ga$.

\Refs \roster \item"[1]" Manin Yu.I., {\it Topics in non--commutative
geometry}. Princeton Univ. Press, Princeton, NJ, 1981.  \item"[2]" Gervais
J.-L., The quantum group structure of 2D gravity and minimal models. Commun.
Math. Phys. 130 (1990) 257-283; Solving the strongly coupled 2D gravity. 1.
Unitary truncation and quantum group structure. Commun. Math. Phys. 138 (1991)
301-338;\newline Alvarez-Gaum\'e L., Gomez C., Sierra G., Hidden quantum group
symmetry in rational conformal field theories. Nucl. Phys. B319 (1989)
155-186;\newline Alekseev A.Yu., Faddeev L.D., Semenov-Tian-Shansky M.A.,
Hidden quantum groups inside Kac--Moody algebras. Commun. Math. Phys. 149
(1992) 335-345. \item"[3]" Karasev M.V., Maslov V.P., {\it Nonlinear Poisson
brackets. Geometry and quantization}. Nauka, Moscow, 1991 [in Russian].
\item"[4]" Shafarevich I.R., {\it Foundations of algebraic geometry}. V.1.,
Nauka, Moscow, 1988 [in Russian]. \item"[5]" Gelfand I.M., Kirillov A.A., Sur
les corps li\'es aux alg\`ebres enveloppantes des alg\`ebres de Lie. Publ.
Math. IHES 31 (1966) 5-20. \item"[6]" Juriev D.V., Complex projective geometry
and quantum projective field theory. Teor.~Ma\-tem.~Fiz. [in Russian]
(submitted). \item"[7]" Drinfeld V.G., Quantum groups. ICM Proceedings,
Berkeley, Calif., 1986;\newline Reshetikhin N.Yu., Takhtadzhyan L.A., Faddeev
L.D., Quantization of Lie algebras and groups. St. Petersburg Math. J. 1
(1990) 193-225. \endroster \endRefs

\enddocument